\documentclass{article}

\usepackage[english]{babel}

\usepackage[letterpaper,top=2cm,bottom=2cm,left=3cm,right=3cm,marginparwidth=1.75cm]{geometry}

\usepackage{amsmath}
\usepackage{graphicx}
\usepackage[colorlinks=true, allcolors=blue]{hyperref}

\usepackage[bottom]{footmisc}
\usepackage{hyperref}

\usepackage{acronym}
\usepackage{csquotes}

\usepackage{authblk}

\title{Less is More: some Computational Principles based on Parcimony, and Limitations of Natural Intelligence}
\author[1]{Laura Cohen}
\author[2,3,4]{Xavier Hinaut}
\author[1]{Lilyana Petrova}
\author[1]{Alexandre Pitti \thanks{Corresponding author: \texttt{alexandre.pitti@ensea.fr}}}
\author[1]{Syd Reynal}
\author[5]{Ichiro Tsuda}

\affil[1]{ETIS laboratory, CNRS UMR8051, CY Cergy-Paris University, ENSEA}
\affil[2]{INRIA Centre of Bordeaux University}
\affil[3]{LaBRI, Bdx. Univ., Bordeaux INP, CNRS UMR 5800}
\affil[4]{Bordeaux University, CNRS, IMN, UMR 5293}
\affil[5]{AIT Center, Sapporo City University, Sapporo, Japan}

\begin{document}
\maketitle

\begin{abstract}
Natural intelligence (NI) consistently achieves more with less. Infants learn language, develop abstract concepts, and acquire sensorimotor skills from sparse data, all within tight neural and energy limits. In contrast, today’s AI relies on virtually unlimited computational power, energy, and data to reach high performance. This paper argues that constraints in NI are paradoxically catalysts for efficiency, adaptability, and creativity.
We first show how limited neural bandwidth promotes concise codes that still capture complex patterns. Spiking neurons, hierarchical structures, and symbolic-like representations emerge naturally from bandwidth constraints, enabling robust generalization. Next, we discuss \enquote{chaotic itinerancy}, illustrating how the brain transits among transient attractors to flexibly retrieve memories and manage uncertainty. We then highlight reservoir computing, where random projections facilitate rapid generalization from small datasets.
Drawing on developmental perspectives, we emphasize how intrinsic motivation, along with responsive social environments, drives infant language learning and discovery of meaning. Such active, embodied processes are largely absent in current AI. Finally, we suggest that adopting \enquote{less is more} principles -- energy constraints, parsimonious architectures, and real-world interaction -- can foster the emergence of more efficient, interpretable, and biologically grounded artificial systems.

\end{abstract}

%%%% Keyword entries to be placed here %%%%
keywords:Energy constraints, Computational Limitations, Efficient Coding, Chaotic Itinerancy, Reservoir Computing, Active Learning, Artificial Intelligence

\section{Introduction}
%\xavier{TODO: fusionner certains paragraphes afin de rendre la structure plus simple. Au max 3/4 paragraphes par page.}
Physical limitations compel Natural Intelligence (NI) to continually surpass itself in order to survive and avoid collapse. It primarily does so by relying on simple yet effective strategies, allowing biological systems to achieve more with fewer resources.
By contrast, modern Artificial Intelligence (AI) faces virtually no restrictions on data volume, time, computational power, or energy consumption, and relies on industrial scale engineering.
While this approach has enabled AI to reach human-level performance in areas like language and cognition, it comes at a considerable cost. Moreover, as novel data becomes scarce, AI’s performance also nears a plateau \cite{jonesAI2024}.

In contrast, the infant brain transcends its inherent limitations, going beyond iterative learning and slow adaptation, despite the limited computational capabilities of its neurons. Its organizational structure supports rapid, flexible development, enabling it to withstand errors and crises. These are the properties of a complex system.
Some of these unique traits of NI are captured by bio-inspired AI, while they are rarely found in classical Machine Learning and generative AI (Transformers).
They also exhibit behaviors that are particularly challenging to replicate, such as abstract thinking, free will, intrinsic motivation, compositionality, structure learning, ad-lib interaction/coupling, and embodiment. These features are central to the core knowledge development of infants in language, planning, geometry, counting, music, and critical thinking \cite{spelke_core_2007}.

%\xavier{
%L’intelligence ce n'est pas ce que l’on sait, mais ce que l’on fait quand on ne sait pas
%Intelligence is not what you know, but what you do when you don't know.

% To rephrase Jean Piaget's citation, NI

Here, we propose that Natural Intelligence (NI) leverages ecological principles to cope with limited resources under uncertain conditions.
To paraphrase Jean Piaget's citation, NI is not defined by what one knows, but rather by what one does when they don't.
Moreover, the constraints on capacity, resources, bandwidth, data, and time naturally underscore the need for efficiency and parsimony.
This perspective on the limitations of intelligence echoes Fields Medal laureate Steve Smale’s designation of \enquote{What are the limits of intelligence, both artificial and human}, one of the key mathematical challenges for the 21st century \cite{smale_math_1998}. %This approach aligns with Smale’s perspective.
This work aligns with Smale’s call to explore those boundaries.
%He identified the limitations of intelligence as Problem 18: ``What are the limits of intelligence, both artificial and human?''

Hence, in this paper, we seek out computational principles that, while remaining as simple as possible, can still capture the complex, efficient, and emergent behaviors underpinning NI and infant cognitive development, from the neuronal level to the organism as a whole.
%
% We choose to present them found at different scales.
% Kahneman, Daniel (2011). Thinking, Fast and Slow. Farrar, Straus and Giroux. ISBN 9780374275631. (Reviewed by Freeman Dyson in The New York Review of Books, December 22, 2011, pp. 40–44)
% Kahneman, Daniel; Sibony, Olivier; Sunstein, Cass R. (2021). Noise: A Flaw in Human Judgment. William Collins. ISBN 9780008308995.
%
% Computational mechanisms at the neuron level are presented in the first part for information representation and encoding using . Those at the neural population level with reservoir computing and chaotic systems are presented in part 2 and 3.
% At the A l'échelle du corps, les mécanismes d'apprentissage actif exploitent les interactions physiques et sociales avec l'environnement comme outil pour accélérer le développement.
%
%The paper is organized as follows. In first sections, we will describe features and results of bio-inspired and dynamical models linked to these cognitive skills such as reservoir computing, spiking neural networks, chaotic neural networks, and methods such as entropy maximization, chaotic itinerancy, intrinsic motivation, with the underlying role of embodiment.
%
At the neuron level, we will start by showing how abstraction can be thought, and represented as low dimensional codes (part 1). Then, at the neural population level, we will present the concepts from dynamical systems of chaotic systems (part 2) and of reservoir computing (part 3). We will see how high-level abstractions can be bootstrapped with little data thanks to random projections as used in reservoir computing: indeed, randomly projecting inputs in high dimensional non-linear dynamics can help to bootstrap generalisation (rules) quickly and thus without the need for numerous data.
We will show how chaos can be beneficial for memory retrieval in high-dimensional space, and for decision making under uncertainties.
%from projecting inputs through time can create complex dynamics from which generalisation can occur quickly, thanks to these high-dimensional representations.
%}
%\laura{
Following this, at the body scale, we will explore the mechanisms of active learning used to leverage the infant mind during physical and social interactions, optimizing convergence under high constraints (part 4).
%}
%
Then, in the discussion, we will address an open scope of the paradox and blessing of limitation. We argue that limitation is an intrinsic component of NI. It is remarkable that computers, digital processing and Information Theory started on the basis of limits of mathematical logic (Turing), computational resources (von Neumann) and physicality of channel communication (Shannon) to produce fantastic results.
Current AI may reintroduce this principle almost forgotten into its roots, and refocus into NI, biology and physics.
%Current AI has forgotten these roots.
% Such leap is required to refocus AI into NI, biology and physics.

%\alex{La nature exploite des mécanismes parcimonieux, et sobres qui suivent une logique d'économie (ou d'éco-logie) pour relever le défi de l'intelligence, à savoir survivre face à l'imprévu, et  être adaptatif en cas d'erreurs ou d'obstacles.
%Bien que ces codes soient simples, ils peuvent produire des compétences ou des comportements complexes, émergents, qui apparaissent de manière non linéaire au cours de l'apprentissage.

% Dans ce papier, nous choisissons  de sélectionner /de présenter quelques-uns de ces mécanismes parcimonieux et efficace que l'on  retrouve à plusieurs échelles.

% Dans la première partie nous présenterons des mécanismes computationels à l'échelle neuronale pour cette efficacité informationelle (ref partie 1), puis des exemples à l'échelle systémique (population neuronale) avec des reservoirs de dynamiques, le reservoir computing (ref partie 2) et les systèmes chaotiques (ref partie 3).
% A l'échelle du corps, les mécanismes d'apprentissage actif exploitent les interactions physiques et sociales avec l'environnement comme outil pour accélérer le développement.

% Dans la seconde partie nous présenterons

% Des codes bruts, aléatoires ou guidés par la nouveauté peuvent avoir des comportements chaotiques, dynamiques, multi-échelles qui peuvent guider l'interaction, l'acquisition le développement et la préservation de la mémoire.}

% First part
\section{Dealing with limited capacity, temporal constraints, and under uncertainty}
% Efficient coding in biology according to Horace Barlow / Alex

% According to F. Alexandre,
The way the human brain performs explicit cognitive processing stems from fundamental limitations\footnote{\url{https://theconversation.com/chatgpt-ma-dit-que-lillusion-de-la-discussion-avec-lia-nous-mene-a-lerreur-238443}}.%\cite{Alexandre2024}.
First, we have a limitation in computing time: we sometimes need to adapt quickly to new situations. To do this, we have developed rapid learning capabilities whereby one or two examples are enough for us to learn continuously, to adapt to a new situation or to retain some of its characteristics.
While the processing time of current AI (Large-Language models) appears very fast, it relies on heavy resources to make it work fast using parallel processing, and the massive learning stage can last for weeks.% can it uses brute force to learn everything off-line
% While the processing time of current AI (Large-Language models ) is very fast, it first requires massive learning requiring billions of words and which can last for weeks.
%This is not the case with Large-Language models (ChatGPT). While its processing time is very fast, it first requires massive learning requiring billions of words and which can last for weeks.
%
Second, we have a limitation in computing power. Our brain has major energy constraints that drastically limit our working memory, that is to say the number of information that we can process at the same time, or the operating frequency of our neurons. While modern computers, ever more efficient and faster, are breaking records in terms of computing power and storage capacity.

%Second, we have a limitation in computing power. We have been living for thousands of years with roughly the same brain, in a world that has become increasingly complex. And this brain has major energy constraints that drastically limit our working memory, that is to say the number of information that we can process at the same time, or the operating frequency of our neurons. While modern computers, ever more efficient and faster, are breaking records in terms of computing power and storage capacity.

The way our cognition has found to deal with this problem is to learn to break down complex problems into simpler subproblems to solve (such as memorizing intermediate steps instead of directly learning a complicated path), which has led us to organize our behavior in time and our knowledge into levels of abstraction.
Current AI does not have this type of constraint, for example synthesizing thousands of texts.

% Third, we have communication problems. While there are techniques for a neural network to simply transfer what it has learned to another network, we do not have the possibility of learning from another human being by connecting directly to their brain. Instead, we have developed ``strategies'' such as language, education or culture that require us to learn to explain ourselves and communicate.

{\bf Efficient coding in the brain.} Accordingly, the human brain must develop efficient strategies in order to transmit, preserve or exploit information continuously, robustly, and despite limitations of its neural material.
Thus, instead of seeing limited computational capabilities as a problem or a constraint for the brain, we can think of it more as a design principle for parsimony and simplicity, much like the Occam razor principle: a heuristic to seek for minimalistic models with the least number of elements.

This view is in line with Horace Barlow who proposed that the human brain exploits efficient codes to maximize information capacity of neurons, despite their intrinsic unreliabilities~\cite{barlow_possible_1961}.
Taken isolatedly, the computational performance of neurons is very poor, but at the population level, their interaction rises their performance at a high level. As proposed by Barlow, some neural mechanisms must exist in the brain to code efficiently information beyond brute force.

From an Information Theory (IT) viewpoint, efficient coding would correspond to maximizing entropy (ME), which is the measure of information quantity~\cite{jaynes_information_1957,jaynes_information_1957-1}. Reversely, maximizing entropy is complementary to minimizing noise or free-energy~\cite{friston_free_2006}. ME means therefore to compress information and to reduce its size in memory space. This is done by breaking down information into smaller bits, learnable by neurons of limited capabilities. By doing so, memories are more compact, faster to be learned, and more robust to be retrieved against catastrophic forgetting.

% It is noteworthy that digital computing, which transformed our world with computers, proceeds similarly by cutting into bits information with low dimensional binary codes.

\subsection{Abstraction as low dimensional codes}

{\bf To choose is to renounce.}
%It is intringuing that
Cognitive skills, such as abstract thinking, knowledge representation, and symbolic processing, are all associated with the idea of information reduction.
For instance, decision making relates to the capacity of inhibiting irrelevant information. Inhibition highlights by contrast the elements that count the most. Moreover, planning is associated with the capability to break down a complex problem into smaller ones (divide and conquer). %cut through, instead of being overwhelmed by .
%
%Besides, knowledge representation use discrete graphes and trees to represent the information structure only.
%
%ThIt provides a comprehensive model, although incomplete, that can be readily used.
%
% Thus, decision making requires to suppress information, instead of being overwhelmed by spurious signals.
In this line, the limited information capacity of the brain may serve as a design constraint to represent the most significant information; that is, the most archetypal ones, the underlying information present in raw signals: its structure.
%
%, through inhibition or by the limitation of their material, neurons

%Decision making, associated in occidental rational thinking, and bayesian inference emphasize strongly the 'positive' act of prediction. Besides, most AI models neglect the 'negative' act of suppressing information as described/found in the Ockam rasor suggest it; which can be the synchronization of wrong/spurious causes with uncorrelated effects.

Information structure is defined as the organizational rules that relate elements from each other, like hierarchies, graphes, networks, patterns. They are independent to the nature of the elements \emph{per se}: e.g., a distinct pattern can represent different data that are fitting it. Information structure corresponds to some of the core knowledge that infants possess early in life, including geometry, logic, causation, music, grammar, or planning~\cite{spelke_what_2003, spelke_core_2007}. % $A$ is before $B$ or $B$ is left $$.
%
% Neurons' information bottleneck may permit to be salient to the information structure the relationship between elements
% Our capacity to relate the most relevant information from each other
%
For instance, the skill of recognizing an object or a face requires to relate the several local cues that compose its contour, and constituting a certain pattern; for example, neonates are capable to follow the facial configuration to perform facial imitation in certain conditions~\cite{meltzoff_imitation_1977}, and to recognize faces even at the fetal stage, before birth~\cite{reid2017}. The studies have showed that they are sensitive to the inverted triangle shape formed by the two eyes and the mouth.

This skill is defined as structure learning~\cite{harlow_formation_1949, braun_structure_2010}, and requires just a few amount of information.
It requires to identify and relate the parameters that characterize the most a scene, forming a compositional representation --e.g., directed lights and shadows, partial occlusion of objects, shape of objects.
It is different from associative learning, which approximates a stastistical model from raw data, at the pixel level. In comparison, structural learning focuses on the learning of a model that links together few variables only, whereas stastistical learning will approximate directly the input. This type of learning can be done possibly through generative probabilistic models, or bayesian inference~\cite{tenenbaum_theory-based_2006, tenenbaum_how_2011, friston_supervised_2023}; for example, to extract rapidly the physical rule that relates force with mass and acceleration.

This capacity of abstraction is often thought as the result of a long cognitive process, acquired at the end of cognitive development. However, some evidences show that infants possess it at birth, which is reverse to the way current AI systems acquire it, after the  learning stage.
Hence, we support Chomsky hypothesis that NI may use a sort of structural framework in the brain that makes possible the human capacity for abstraction, starting at birth.

%analyze what we see by finding the good parameters that characterize it.

% Thus, we propose that the limited information capacity of neurons may help them to recognize rapidly information structure in signals.
%
% Dehaene and colleagues propose that symbolic processing is proposed to be a unique trait of

{\bf Symbolic processing, hierarchical codes.} In this line, knowledge representation and symbolic AI have been constructed on data structure, like directed graphes, and binary trees. This approach has been criticized by the connexionist approaches due to the lack of realism, being too far from the way data are represented and computed in the brain. However, current AI use massive neuronal architectures, huge volume of data, and brute force computation that are not compatible with NI.
Paradoxally, the first generation of AI systems, based on symbol processing, like A-star, minimax, Prolog, provide parcimonous and simple solutions in comparison to deep learning, reducing problems into low dimensions. %A similar argument can be opposed to it.
We hypothesize that the poor computational capabilites of biological neurons (spiking neurons) may ideally combine the best of the two approaches to represent data and problems in low dimension, as discrete codes. %The brain may turn into an advantage its poor computational capabilities with simple mechanisms.

{\it Hierarchical Trees.} Among the discrete and low dimension codes that exist, the binary codes are the most used in computer science. Binary codes are also exploited widely in cognitive science to describe hierarchical patterns in language, music, logical rules and action planning~\cite{friederici_brain_2011, dehaene_neural_2015}.
Hierarchical trees are used to represent in compact format the knowledge structure and the relationships among persons (e.g., family trees), and items (e.g., object categories, spatial locations, or actions).
In this line, Neil Burgess proposes that the hippocampus uses the so-called factorization codes to compress the information structure that is present in data~\cite{whittington_tolman-eichenbaum_2020}.
Dehaene and colleagues have a similar viewpoint for neurons in the prefrontal cortex, which have been found sensitive to complexity depth, representing information as nested trees~\cite{dehaene_neural_2015}.
Other experiments in language processing support this observation of the neurons's sensitivity to serial order, permutation and complexity level~\cite{wang_representation_2019}. He further proposes that this feature is at the origin of symbols and concepts, which makes human intelligence unique with respect to other species~\cite{dehaene_symbols_2022}.

In comparison, recent works found no evidence of formal reasoning in large language models, but limited capacities to learn symbols and rules~\cite{lakretz_can_2022, mirzadeh_gsm-symbolic_2024}.
Instead, LLMs likely perform a form of probabilistic pattern-matching and searching to find closest seen data during training without proper understanding of concepts.
By doing so, their behavior is better explained by sophisticated pattern matching.

{\it Serial Order.} Other popular low dimensional codes are the serial order codes, which represent information as directed graphes.
The problem of serial order is ubiquitous in perception, memory, and action~\cite{rosenbaum_problem_2007}.
According to Lashley, the serial ordering of actions and its recall impose a non-commutativity in the representation of actions within the brain to coordinate elements accurately in time~\cite{fitch_hierarchical_2014}. Ordering in time and space events requires to abstract them as chunks, and therefore to represent them as symbols, to manipulate them as if they were discrete items~\cite{pitti_search_2022}. %representation for prediction and anticipation.

{\bf Infant's core knowledge development.}
Tenenbaum and colleagues emphasize the role of abstraction for logic and prediction in infants~\cite{tenenbaum_how_2011}. Abstraction can be associated with relational thinking, in the form of bayesian graphes and networks~\cite{pearl_causality_2009, pearl_book_2018}.
%
%In comparison to ML,
In developmental science, Baillargeon highlights that infants learn at a very high pace the information structure presents in the world~\cite{baillargeon_physical_1994, baillargeon_core_2012}. Counter-intuitively, she proposed that infants analyze a situation much more at an abstract level rather than through the senses. For instance, Baillargeon studied the acquisition of the concepts of object occlusion and of object permanence, and identified their capacity to relate few elements in a scene, temporally and spatially. In line with it, Spelke identified how infants make hypothesis in the world, like little scientists, using bayesian causal relationships~\cite{spelke_what_2003, spelke_core_2007}.% to construct a high-order rule. This rule can be then tested, assessed, or forgotten during life.

Causal reasoning in space and time through embodied experiences is therefore highly related to abstract thinking. For instance, before being formalized into an equation, the Newton law $F=m.a$ is implicitly experienced through the bodily senses by relating internal forces with the external masses of weighted objects and their relative speed. This overt physical rule is sensed and tested rapidly by the many sensorimotor experiences to assess its validity. An amodal model is learned to extrapolate predictions in new situations, and with novel objects.
%
% The relationship is causal and bidirectional.

Spatio-temporal ordering is also an important element in planning, perception, and reasoning~\cite{pitti_gated_2020, pitti_search_2022}. For instance, directional graphes and serial order codes can reduce the dimensionality of the problem space of plateau games and their combinatorics, like the Hanoi Tower game~\cite{wand_serial_2024}. The objectives of the Hanoi Tower game is to place distinctive palets in a specific spatial order. While the combinatorics of possible states and sequences augment by an exponential scale in $2^{N-1}$, in which N is the number of palets, the number of rules rests circumbscribed to a low dimension.

This mathematical game, like other plateau games, imposes the learning of a model.
For instance, adding more palets increase the game complexity but does not change the game: the rules are the same and we can still play to it and compose new solutions. However, compositionality is still difficult for current AI systems as they cannot easily compose sequences from unseen states.
As another example, information structure in visual experiments, like force contact or 3D mental rotations are often tackled as an inference problem in sensorimotor space by generative AI systems. As such, they rely heavily on large image database to approximate a model, still with difficulties to understand things like gravity or human morphology. In constrast, algebraic patterns found in physical rules and information structure are small, and can generate infinite number of solutions.
Algebraic and geometric problems can be reduced to low dimension problems based on generic rules. The same is true for language~\cite{dehaene_symbols_2022}.
% For instance, when manipulating a dice, of sensorimotor coordination.
% For instance, although the number of observed rotations of one solid object can be infinite, the geometrical rules of it are limited. The relative spatial information move together and can be represented as a whole.

% comparison between generative [dehaene].
%The Necker Cube illusion is based on a bistable representation of two consistent partial views. Taken as a whole, these relative representations conflict from each other at the conscious level, and only one consistent representation is perceived at a time.

% {\bf [language and counting]}

{\bf Neurons' sensitivity to spike order.}
Within the brain, biological neurons have been found sensitive to the serial order of the spikes~\cite{thorpe_spike-based_2001,izhikevich_spike-timing_2004,izhikevich_polychronization_2006}.
%. In [],
Bi and Poo discovered a dopaminergic mechanism based on reinforcement learning between post-synaptic and pre-synaptic neurons, and called spike timing-dependent plasticity (STDP)~\cite{bi_activity-induced_1998}. This hebbian learning mechanism relies crucially on the temporal dimension in spike trains.

In terms of codes, Perceptrons embed high dimensional information with synaptic weights of high precision (e.g., 64 bits), which is convenient to embed rich signals with $R_{W perceptron}=R_X$, and $R_X \in R_e$.
In comparison, spiking neurons possess limited computational capabilites, and work in very low dimension and resolution. The serial order codes of spiking neurons, sensitive to the temporal order of the spikes, are in dimension $R_{W spike}=L$, with $L$ the number of pre-synaptic neurons connected; with $L \in N$ such that $L \ll R_X$. Spiking neurons provide a form of computational efficiency as they provide low resolution codes, as the binary code~\cite{pitti_digital_2022, pitti_moreless_2024}.

Observations in the Broca area and in the pre-Supplementary Motor Area (pre-SMA) located in the Pre-Frontal Cortex (PFC) confirmed the existence of neurons sensitive to ordinal and hierarchical patterns only; patterns like ABAB, or ABCABC, independently to what item is in A, B, or C. For instance, some neurons were found sensitive to the temporal order in audio sequences and to proto-grammars (i.e., a rudimentary grammar) but not to the particular sound emitted~\cite{friederici_brain_2011, benavides-varela_learning_2017, gervain_neonate_2008}: e.g., ``tomato'', ``kabuka'' but not ``totomi''. Other neurons were found sensitive to the different levels of complexity and depth in sequences~\cite{petersson_artificial_2012, fitch_artificial_2012}. In the motor domain, a majority of neurons in PFC were found sensitive to the temporal patterns
of motor behavior such AABB or ABAB and a minority to the particular motor units~\cite{tanji_behavioral_2001, shima_categorization_2007, tanji_concept-based_2007}. Moreover, different ones were observed salient to the temporal coherence in visual scenes~\cite{fadiga_brocas_2009}; e.g., its semantic.
This class of neurons was particularly studied twenty years ago as they were found to trigger not only to one's own actions but also to those of others. They correspond to the so-called mirror neurons system in the monkey's frontal cortex~\cite{rizzolatti_neurophysiological_2001}.

Some similar results were found with neurons sensitive to orders, schemata in spatial contexts in PFC~\cite{barone_prefrontal_1989}, and to geometrical rules in the recognition of shapes like triangles or rectangles~\cite{averbeck_neural_2003}, or in visual sequences~\cite{dehaene_neural_2015}. Surprisingly, these neurons were all found insensitive to the particular sound, action or visual information composing the sequence presented.

{\bf Efficient coding by information suppression.}
From an IT viewpoint,
this capacity of abstracting a model from raw signals can be assimilated to an information reduction process or to a compression. % or to redundancy suppression
Restricting the information capacity of neurons $R_W$ impose to find the most comprehensive way to represent data complexity $R_X$.
Entropy maximization can serve then as a reinforcement signal to reduce the dimensionality of information~\cite{pitti_digital_2022}.
Using the source coding theorem of Shannon, we can relate then the two variables such that we have $\log R_X \approx k \log R_W$, with $k$ the number of low dimensional neurons necessary to approximate the high dimensional input $X$.
EM may represent then the internal drive that reduce the complexity of the world ($R_{X}$) into a low dimension model ($R_W$), in brains and bodies~\cite{pitti_informational_2024}.
The brain may benefit from the restricted computational capabilites of spiking neurons, to work in low dimension, with minimum information.

\subsection{Chaotic itinerancy and information maintenance over long time scales}

{\bf How to Go Beyond Static Artificial Intelligence.}
In this subsection, we discuss the dynamic association of memories (DAM), which is considered one of the origins of natural intelligence. In deep neural networks and their extensions, such as transformers, a single association of memories is already embedded~\cite{hopfield-net}, enabling the system to carry out preprogrammed or newly learned actions. For example, ChatGPT can associate word B with word A based on learned probabilities, which are contextually determined through a massive learning process. As von Neumann pointed out~\cite{neumann}, it is not implausible to think that human natural languages can emerge probabilistically in various contexts, especially considering the diversity of languages. ChatGPT, for instance, can learn and generate human languages to a degree that satisfies human communication needs.
However, natural intelligence, as represented by biological neural networks, is fundamentally different, even when considering language. Current artificial intelligence (AI) systems do not genuinely understand the meaning of the languages they process. The association of memories in these AI systems relies on neural states with multiple attractors, resulting from frustrated interactions, much like the spin-glass model~\cite{hopfield}.

There are more realistic, albeit still abstract, neural models (for example, see~\cite{Hopfield-bio}). Among these, we focus on nonequilibrium neural networks with inhibitory neurons, which play a role in the self-organization of networks exhibiting DAM~\cite{tsuda87,tsuda92,tsuda01}. Reflecting the structure of biological neural networks in the neocortex~\cite{szentagothai} and hippocampus~\cite{buzaki}, the inhibitory neurons in DAM models do not consume memory but temporarily mask attended memories (discussed further below). Sensory inputs are received in the limbic system at an early stage of sensory information processing, then further processed in the neocortex, creating a global attractor across the brain that expresses awareness, aided by intention~\cite{freeman-book}. This masking of attended memories drives the successive association of memories, possibly hinting at the emergence of conscious thought. Biological neural networks can learn from stimuli and relearn their internal states, even during the masking process~\cite{tsuda92}, a process we call compensation learning~\cite{Koerner}.

{\bf Attractor Dynamics and Beyond.}
Each memory can be represented by an attractor with a specific range of attraction in phase space, known as the basin of attraction. It is well known that spurious modes, or spurious attractors, can also emerge in associative networks unless memories are orthogonalized. When inhibitory neurons mask a spurious attractor after a trajectory is drawn to it, a sequence of spurious attractors forms dynamic trajectories, potentially chaotic due to the sensitive dependence on frustrated states. This overall dynamics, known as chaotic itinerancy~\cite{kk, ki, it-scolar, kk-it, it-curr.opin.}, manifests as a sequence of temporary convergences to memory attractors, interspersed with chaotic trajectories of spurious attractors.

The structure of memory space can be represented as follows: let $P^i(k)$ be a memory state $i$ at time $k$, where $i=1, \cdots, m$, and $\Psi(P^i(k), \cdots, P^m(k))$ be a memory space at time $k$. Then, masking a memory state $j$ at time $k-1$, $P^j(k-1)$, is expressed as $\Psi(P^i(k), \cdots, P^m(k))-\Psi(P^j(k-1))$. It is important to note that this masking process differs from $\Psi(P^i(k), \cdots, P^j(k)-P^j(k-1), \cdots, P^m(k))$, which represents unlearning the memory $j$ by modifying the weight matrix~\cite{crick, hop-unl}. Masking begins with the activation of intended and attended activities in the limbic system and neocortex~\cite{freeman-book}.
Masked memory attractors become quasi-attractors with neutral stability, making them different from conventional geometric attractors. During masking, memory states exhibit both stable and unstable manifolds in a different way from saddles. These quasi-attractors may be mathematically described as Milnor attractors~\cite{milnor}, where the first derivatives of vector fields vanish, leaving higher-order stability to dominate. This leads to tangencies between multiple basins of attraction, enabling transitions between memory states~\cite{kaneko-milnor}.

We observe that recurrent neural networks (RNN), formed by cortico-cortical interactions through pyramidal cells, and the arborization of these cells within columns, are associated with inhibitory neurons such as Martinotti and chandelier cells. These cells inhibit pyramidal cells at the dendrites and the axon initial segment, respectively, receiving signals related to attended memories. We demonstrated the successive association of attended memories via the masking of currently attended memories in both probabilistic control models~\cite{tsuda87,tsuda92,tsuda01} and deterministic inhibitory control models~\cite{tsukada-tsuda}. If the prefrontal cortex monitors this process of chaotic itinerancy, it may define self-consciousness, as suggested by Dehaene's definition~\cite{dehaene}.

{\bf Superiority of Chaotic Itinerancy: Its Relation to Natural Intelligence.}
Chaotic itinerancy arises from weak chaos, where many Lyapunov exponents converge near zero, indicating neutral stability, and even the largest exponent remains small. Weak chaos plays a crucial role in information processing since the quantity of input information or initial conditions does not rapidly decay. Thus, the network maintains information dynamically~\cite{mt85,mt87,mt89}. This weak chaos influences both neuronal differentiation in evolution and functional differentiation in the development of neural systems. We demonstrated that these natural intelligent processes can be realized in evolution models of dynamical systems networks~\cite{wata-t} and also in extended reservoir computers~\cite{yy-t,t-entropy}.

The present proposition, based on the neural network's ability to generate chaotic itinerancy, outlines how natural intelligence emerges through associations with emotion and intention. This is markedly different from AI’s reliance on massive data and computations, which demand enormous energy resources. In contrast, the proposed neural learning is highly efficient, requiring only a few hundred to a maximum of 0.1 million neurons and relatively short learning times, such as a few hours. This efficiency could be one of the key elements of natural intelligence in the human brain. Furthermore, the proposition has practical applications, such as developing intelligent robots capable of resilient and flexible responses to environmental changes~\cite{kawai, pitti, yano, kozma}, facilitated by chaotic itinerancy~\cite{kuniyoshi1, kuniyoshi-nakajima}.

\subsection{Reservoir Computing: Bootstrapping abstraction with complex dynamics}

% TODO later
%\xavier{voir par rapport à Gerschmann 2017 comment le réservoir computing se distingue: http://dx.doi.org/10.1080/17470218.2016.1159706}

%\xavier{DONE: conseil d'alex: enlever la partie historique du début}

\textbf{Reservoir Computing emergence.}
It is often stated that Reservoir Computing (RC) has emerged twice in 1995~\cite{buonomano1995temporal,dominey1995complex} from the computational neuroscience side, although it can be argued that similar forms have appeared previously several times (see the references collected by Herbert Jaeger on Scholarpedia\footnote{\url{http://www.scholarpedia.org/article/Echo_state_network}}~\cite{Jaeger2007scholarpedia}).
%Thus, it appeared only some years after the famous {\it SRN} from Elmann in 1990, which was itself featured a few years after the invention of {\it BPTT}~\cite{werbos1988generalization,werbos1990backpropagation}. Thus, Reservoir Computing can be seen as a possible \enquote{end of the road} of simplification of \emph{RNN} training: first RNNs were fully trained with BPTT, then only one step back in time of BPTT is performed with SRNs, and finally inputs and recurrent weights are not learnt anymore with the RC paradigm.
RC has emerged as the form we know today in early 2000's with the Echo-State Networks (ESN) of Jaeger~\cite{jaeger2001echo} and with the LSM of Wolfgang Maass and colleagues~\cite{maass2002real}.
%A RC community started to take shape: machine learning community was more focused on ESNs and computational neuroscience more on LSMs~\footnote{Even if ESNs or equivalent (rate-coded RNNs) were also used in computational neuroscience, e.g. }. This movement was probably enhanced because of the nice performances obtained by Jaeger on chaotic time series prediction~\cite{Jaeger2004science}.

\textbf{Biology.} Reservoir Computing (RC) emerged at start from the computational neuroscience side~\cite{buonomano1995temporal,dominey1995complex,dominey1995model,maass2002real}, before emerging also in the machine learning side~\cite{jaeger2001echo,jaeger2002adaptive,Jaeger2004science}.
Indeed, a reservoir can be seen as a canonical computation unit~\cite{haeusler2007statistical}; it could model a cortical column: what computational neuroscientists often consider as a generic unit of computation.
Since 1995,  Dominey have used it to model the cortico-basal network~\cite{dominey1995complex}: the reservoir playing the role of the (prefrontal) cortex and the output layer playing the role of the striatum (input of the basal ganglia from the cortex).
Dominey showed that even with random networks (that were not called reservoirs yet) it was possible to observe similar neuronal activation patterns~\cite{dominey1995model} then in studies on sequence processing in monkey prefrontal cortex~\cite{barone1989prefrontal}.
RC developed much faster in the machine learning community since the 2000's, but in the 2010's it became more popular from the experimental neuroscientists side. Neuroscientists started using this idea of high-dimensional non-linear representations that can be decoded by a linear classifier. It was a new way to interpret electrophysiological recordings from monkeys~\cite{machens2010functional,rigotti2013nature,enel2016reservoir}: the idea was no longer to find particular sequential pattern in neural activity (like in ~\cite{barone1989prefrontal}), but rather to just decode linearly if some information were present.

\textbf{Intuition.}
Reservoir Computing is a paradigm to train Recurrent Neural Networks (RNN) without training all connections.
The names ``reservoir'' for the recurrent layer, and ``read-out'' for the output layer, come from the fact that a lot of input combinations are made inside the recurrent layer (thanks to random projections). The reservoir is literally a reservoir of calculations (= ``reservoir computing'') that are non-linear.
From this ``reservoir'' one linearly decodes (= ``reads-out'') the combinations that will be useful for the task to be solved.
Reservoirs can be implemented on various kinds of physical substrates~\cite{tanaka2019recent} (e.g. electronic, photonic, mechanical RC).
In other words, the reservoir is a random RNN which already produces \enquote{natural} dynamics\footnote{It is important to note that these dynamics are deterministic, even if the network is generated with random weights (as soon as no noise is injected in the reservoir states).} given a flow of inputs. Then the \enquote{richness} of these dynamics are used through the linear read-out layer to predict/classify. This means that these natural dynamics -- and thus, inputs and recurrent weights --  don't need to be trained, which represents an interesting advantage in computational cost.

\textbf{The kernel trick.}
An intuitive way to understand how reservoir computing works is to think it as a temporal Support Vector Machine (SVM)~\cite{verstraeten2009reservoir}.
Suppose you want to separate blue dots from red dots, but in your initial 2D space you cannot separate them with a line.
With a SVM~\cite{vapnik1999nature} you project theses inputs (i.e. the dots) into a higher dimensional space.
In this high dimensional space you can find a hyperplane (an equivalent of a \emph{line} in higher dimensions) that separates your blue dots from your red dots. Finding this hyperplane is equivalent to perform a \emph{linear} regression.
You can have different types of kernel with an SVM; in reservoirs this kernel is random.

%\xavier{dire que le reservoir computing c'est similaire à des méthodes évolutionnaires
%où il y a un "échantillonage" avec une population d'individus puis une "sélection" des meilleurs individus. le reservoir computing c'est similaire car on projette en grande dimension plein de combinaisons des entrées (= un grand échantillon de nombreux calculs) puis avec la régression linéaire on vient sélectionner (avec pondération) les calculs qui sont les plus utiles pour les sorties que l'on veut obtenir. Cette idée peut aussi être liée à l'idée d'exploration et exploitation: on explore avec le réservoir de calculs, puis ont exploite les calculs qui sont utiles.}

\textbf{Sampling, Selection, Combination.}
We can make a parallel with other methods using randomness. In evolutionary methods, such as genetic algorithms, a population of individuals (i.e. candidate solutions to a problem) is generated and then some individuals are selected, before being combined (e.g. by cross-over and mutation) to generate the next round population. At each round, the newly created sample of candidate solutions is evaluated and then selected for the next round. We could see reservoir processes in a similar fashion. First, a lot of inputs combination -- together with past reservoir states -- are performed with  random projections (i.e. sampling). Second, the read-out selects the useful combinations available in the reservoir to combine them to produce the desired output (i.e. selection and combination). Such view puts into perspective the more \enquote{natural} (i.e. biological) side of reservoir paradigm compared to back-propagation paradigm.
We could similarly see this two steps reservoir approach as exploration and exploitation as in the Reinforcement Learning paradigm: a lot of combination of inputs and past reservoir states is explored, and then they are exploited by the read-out layer.

%\xavier{enlever cette partie sur robot une fois que abstraction est fait}
%\textbf{Robot Language Acquisition Modelling via Cross-Situational Learning with Little Data.}
%
%How do children bootstrap language through noisy supervision? Most prior works focused on tracking co-occurrences between individual words and referents. We model cross-situational learning (CSL) at sentence level with few (1000) training examples. We compare two recurrent neural network architectures often use as cognitive models: reservoir computing (RC) and LSTMs on three datasets including complex robotic commands. Surprisingly, reservoirs demonstrate robust generalization when increasing vocabulary size: the error grows slowly compared to an LSTM of fixed size. This suggests that that random projections used in RC helps to bootstrap generalization quickly. How robots acquire basics of language like in child-caregiver (Human-Human) interactions could give hints of how to link animal vocalisations with behaviour in ambiguous context. Cross-statistics between sequence of vocalisations and various contexts could probably be learnt in few trials by such Reservoir architecture.

%\xavier{ECRIRE PARTIE SUR boostrap abstraction with complex dynamics}
%\xavier{xav: -> reprendre mes notes de prés sur bootstrap abstraction dans ma présentation de ICDL 2024 et ce que j'ai repris après cette présentation}

\textbf{Bootstrapping abstraction.}
In order to understand human quick ability to learn language with little data compared to Large Language Models, we performed experiments on the modelling of language acquisition with RNNs, as if a child or model embedded in a robot would have to learn the meaning of simple scene description just by being exposed to co-occurrences of scenes and sentences describing part of the scene~\cite{juven2020cross, variengien2020journey, oota2022cross}. The task is not so easy for RNNs because (1) sentences describe only part of the scene, and the model don't know the meaning of any word at start; and because (2) we only use 1000 sentences to train the models, which is small due to the number of combinations of objects, colors and positions that can occur. In order to create different levels of difficulty we generated corpora of different complexity, the easiest containing only 4 different objects and the harder one containing 50 different objects (without changing the training size). We compared two recurrent neural network architectures: Reservoir Computing (RC) and LSTMs. Surprisingly, reservoirs demonstrate robust generalization when increasing the vocabulary size: the error grows slowly compared to an LSTM of fixed size.
In~\cite{variengien2020journey}, we made several analyses the internal activation of reservoirs and LSTMs. We showed that LSTMs learn to create relatively sparse representations, in particular internal unit activity change for few words only, as some kind of \enquote{mixed selectivity}~\cite{rigotti2013nature}. On the contrary, reservoir activity showed a highly diverse and distributed activity. In a context where one agent has to learn a task from scratch with a small dataset, it seems that the reservoir representations based on random projections are very useful. This suggests that these random projections help to bootstrap generalization quickly.
This seems to be confirmed by two other recent experiments. First, we showed that adding a reservoir for preprocessing the inputs given to a state-of-the-art reinforcement learning algorithm could reduce the training time (which is usually in millions of steps)~\cite{leger2024evolving}. Second, in a prediction of COVID-19 hospitalization with only 400 days of data with 400 features, RC is the best performing method compared several others~\cite{ferte2024reservoir}. Thus, RC seems to have the ability to create good abstraction of inputs with little data.

\textbf{Random projections}
Random projections are know for their intriguing general properties, especially as a practical method for reducing the dimensionality of a set of vectors while maintaining its structure.
The Johnson–Lindenstrauss lemma states that  high-dimensional data could be projected into a low-dimensional feature space with nearly negligible information loss~\cite{johnson1984extensions}.
It has important application in various domains and for instance in data compression.
For instance, in \emph{Compressive Sensing} a signal having a sparse representation in one basis can be recovered from a small number of projections onto a second basis that is incoherent with the first, and random vectors  provide a useful universal measurement basis that is incoherent with any given basis with high probability~\cite{duarte2006universal}.
In summary, random projections are useful generic operations that preserves distances between vectors and have practical applications in various fields; as they come nearly \enquote{for free}, biological systems are likely to exploit this principle.

\textbf{Reservoirs of computations}
Reservoir Computing leverages interesting properties of random projections. However, we still lack precise mathematical explanations for the practical success of RC, which suggests that there may be other properties of random projections yet to be discovered and demonstrated.
Beyond random projections, Reservoir Computing can exploit a virtually unlimited number of physical computing media \cite{tanaka2019recent}. Indeed, it is possible to use the properties of different physical systems, such as light, magnetic properties of superconducting materials, or even biological processes, to create reservoirs. What can we conclude from this? Equivalents of random projections are probably available in large quantities across all the physical, chemical, and biological systems that surround us. Pragmatically, Reservoir Computing has the potential to become a far more eco-friendly computing paradigm than the one based on gradient backpropagation.
Just as a windmill uses wind to grind grain or generate electricity, Reservoir Computing exploits natural dynamics to perform computations. This paradigm has the incredible advantage of working with a wide range of systems, whether they are composed of random neural networks or not. Pragmatically, a windmill needs to be placed in a location where the wind is sufficient. Similarly, for each task, it suffices to find a reservoir with natural dynamics that are well-suited.

Extending the windmill analogy, we propose a roadmap to encourage broader adoption of Reservoir Computing. Just as weather maps are used to determine locations where the wind is often strong, it would be useful to create similar mappings for Reservoir Computing. Each location on this \enquote{Reservoir Map} would correspond to hyperparameter (HP) coordinates indicating the types of reservoir dynamics observed in that configuration of HP. For each task already known in the RC domain, such a mapping could be constructed, leading to a predictive map for future tasks. For a new task, one would then identify which existing tasks it is equivalent to, thus predicting the regions of the Reservoir Map where this new task could find reservoir dynamics that work well. This map could not only be created for \enquote{classical} reservoirs (based on artificial neural networks) but also for different physical reservoirs. For each task, we could then determine which physical media provide the most interesting natural dynamics (and hence the best performance). This could encourage a broader adoption of reservoirs, including physical ones, in applied research in the AI industry.

\subsection{Active learning and social interaction for scaffolding}
Babies are intrinsically limited, on the sensors, actuators and computation levels. However, those limits can actually contribute to their capacity to learn faster and more efficiently than AI. In this section, we adopt a developmental perspective to discuss the importance of intrinsic motivation coupled with social interaction for artificial agents to develop the capacity of meaning.
We illustrate the importance of 1. intrinsic motivation and 2. social interaction between the caregiver and the infant in the context of the acquisition of communication in humans and robotic agents. We posit that both these aspects are key to our capacity to make sense of the world and to acquire a meaningful language, while they remain rarely exploited in current AI systems.

{\bf Active learning during infant's language development.}
%\subsubsection{Active learning and intrinsic motivation during infant development}
Current approaches to infant learning focus mostly on a passive view of language learning, as the agent is exposed to a huge quantity of data that it must learn to replicate (e.g., current Large Language Models). In this classical view, learning is equivalent to forming statistical association by listening to a caregiver who masters the language already, or being exposed to a large quantity of text. In contrast, we posit that agents must be active to make sense of the world. The classical concept of learning meanings consists in the acquisition of associations between arbitrary symbols (the words) and referents in the real world (the objects). The capacity of infants to learn this association so efficiently has been explained with cross-situation statistical learning \cite{smith2008infants}. For instance, this paradigm was recently applied to an AI system to learn word-referent association from a camera mounted on a baby's head \cite{vong2024grounded}. %For example, in [], the authors propose to model infants language learning by using videos from a visual input coupled with an audio stream acquired from a camera mounted on the head of a baby. They rely on cross-situational learning to explain how babies can solve the word-referent problem, i.e., the fact that when they hear a word in presence of many objects, the actual referent is ambiguous.
 Although literature in developmental psychology shows that statistical learning and contingencies detection between words they hear and the surrounding environment are essential for language development, it also emphasize the key proactive role of the infant \cite{golinkoff2019language}. In \cite{tamis2018taking}, the authors propose to \enquote{flip the lens} to consider infant behaviors and changing skills across development as primary catalysts for learning language. Another argument to consider the active role of the learner is the fact that infants are able to express meanings well before they produce their first words. For example, babbling infants use vocalizations to request objects and actions from their caregiver. During this phase, their verbal utterances are not yet imitations of adults' words \cite{halliday2004language}. This suggests an active and creative role of the learner in the acquisition of language. In this view, the learner actively tries to achieve specific goals by using language as a tool. This functionalism of language is defined by Bates as the idea that the form of natural language is created, acquired and used in the service of functions \cite{halliday2004language}\cite{bates1991functionalist}. We posit that this active and functionalist view of language learning can also contribute to learn more effectively for artificial agents.
 %From birth, and even earlier, infants possess the ability to learn language. They can identify phonological, semantic, and grammatical patterns from the language they hear, and quickly recognize the contingencies between words and their surrounding environment, which are key skills essential for language development.

{\bf An approach based on intrinsic motivation for language acquisition.}
\cite{cohen2018social} proposes an alternative solution to state of the art approaches of symbolic words acquisition, based on a functionalist view of language. In that approach, basic intrinsic motivations, such as thirst, hunger and boredom constitute the base to develop the first steps of language in young infants. This perspective makes it possible to express language learning as a reinforcement learning problem, in which the capacity of the infant to be understood by his caregiver leads to obtaining a desired object. Thus, only understandable words are reinforced. The results show that it is possible for the agent to learn mutually understandable babbling words to refer to objects, even before the baby is able to use the adults' language. Furthermore, the authors observe an emerging expansion of the diversity of the speech used by the baby, as babbling utterances that are too similar to refer to different objects are less understandable by the caregiver.

%\iffalse
%\begin{figure}[htbp]
%    \centering
%    \includegraphics[width=0.5\textwidth]{"babbling.png"}
%    \caption{Your caption here}
%    \label{fig:babbling}
%\end{figure}
%\fi

{\bf The role of the social environment and caregiver responsiveness to learn to communicate}
Another key property of infants learning consists in their capacity to leverage their social environment to help them learn faster. For instance, they can rely on their caregiver's guidance, reaching initially competences in their zone of proximal development: in the presence of a caregiver, infants can achieve competences that they are not yet able to master alone \cite{vygotsky1986thought}. This aspect is well illustrated by language acquisition, that depends deeply on the social environment and on the caregiver's behaviors \cite{tamis2014infant}. For example, caregivers adapt the difficulty of the task to the current competence of the infant. In \cite{cohen2018social}, the authors model such a scaffolding behavior of the caregiver in the context of an object reference task of a pointing gesture. For this purpose, the caregiver initially accepts vague pointing toward the target object, but decreases his tolerance at each success of the infant. Results shows that convergence is reached faster when the caregiver adopts such a strategy, as opposed to a caregiver that has a low tolerance to mistake from the beginning of the learning phase.
In \cite{lemhaouri2022role}, the authors further investigate the influence of affective and social context by examining two key properties of caregiver behavior: the contingency and the contiguity of the caregiver's responses to the robot. Contingent responses are those that are conceptually dependent on the child’s exploratory and communicative actions, while contiguity refers to the temporal alignment of the caregiver’s responses with the child’s behavior. The findings demonstrate that these two factors are central to language learning, aligning with developmental research showing that parental responsiveness is a strong predictor of language milestone acquisition \cite{tamis2014infant}.
Another work from the same line of research explores further the active and social role of the robot in language acquisition by studying the role of social feedback. In \cite{markelius2023human}, a Differential Outcomes Training (DOT) protocol is proposed whereby the robot provides feedback specific (differential) to its internal needs (e.g. ‘hunger’) when satisfied by the correct stimulus (e.g. cookie). Results show that robot’s language acquisition achieves faster convergence in the DOT condition compared to the non-DOT control condition.

\section{Discussion}

%\xavier{j'ai inversé l'odre de ces deux parties}

%\xavier{Syd PART}

%Syd
\paragraph{A renewed vision of what could be asked of A(G)I: limits as an optimization engine}

Just as we rarely question the aims of natural intelligence, we do not question those of artificial intelligence either. Stated differently, while we frequently reflect on the perimeter of A(G)I, we less often ponder on what we do expect of it, e.g., from a societal and historical perspective, both explicitly (in public statements) and implicitly (as fantasies). It is indeed generally assumed that world-models fulfil a predictive purpose~\cite{Schmidhuber2010}. From the point of view of the survival of the species, natural intelligence also fulfils this function, since the ability to predict enables us to anticipate dangers and optimise access to resources, and one might be tempted to say that those individuals who have not developed this intelligence have not had descendants up to the present day~\cite{Dawkins06}.

We can hypothesise that the optimized computational functioning - with \enquote{less} resources - that the human brain has implemented is the result of the conjunction of two factors: on the one hand, the existence of strong energy constraints with significant consequences on computing power ; on the other, the survival instinct that has driven members of the species to quickly find efficient solutions (those guaranteeing survival) while satisfying the aforementioned energy constraints~\cite{Friston2013}.
%\alex{je pensais comme refs What is Life de Schrodinger, Karl Friston sur la free energy}.
In other words, natural intelligence of individuals has effectively contributed to: i) finding solutions to practical situations involving survival; ii) yet also, improving cognitive functioning  in order to optimize that same natural intelligence under energy constraints, which indirectly contributed to the effectiveness of the first item. It is thus the very existence of limits that has led individuals to improve their cognitive abilities. More specifically, there have always been two complementary types of limits at play here: a limit set by energy constraints along with the limit to the human lifespan. While the former poses the usual constraint framework akin to any constrained optimization problem, it is the second limit that embodies the impulse that drives individuals to carry out any survival-oriented optimization scheme, %\xavier{to optimize what?} \ syd{dans un sens tres general: tout processus d'optimisation visant a la survie}
leaving them little choice indeed since, in the absence of this will, these same individuals would die quickly. From a purely epistemological point of view, what we have here is distinctly a self-making cybernetic system driven by survival constraints, or values-driven — provided we adopt the perspective that the survival of individuals is now a fundamental value of our societies, whereas it used to be a mere constraint for the transmission of \enquote{survival capacities} to descendants.

Taking into account the complex interaction of these two limits is anything but trivial in terms of the paradigm shift it can bring about. It helps indeed shifting the focus from an heteronomous % \alex{AGI? expliquer heteronomous: par ex. 'heteronomy being the opposite of autonomy'}. \Syd{done}
optimisation of cognitive functioning, to that of A(G)I autonomy, where by heteronomous optimization we mean an artificial optimization process that is governed by external forces, rules or laws. World-models may describe the world in an extremely comprehensive way, and as such offer every guarantee of reliability as a process for predicting the future: in their current form they remain systems subject exclusively to heteronomous laws.
As such, autonomy, in this operating regime, would result from the existence of this dual system of limits, one of the consequences of which would be a form of \textit{artificial} natural selection.

Current AI systems are not anthropologically symmetric with respect to humans: they are not autonomous, but merely heteronomous\cite{Latour1994}.
What we call autonomous agents, such as autonomous vehicles, remain in fact heteronomous. The symmetrisation of the anthropological relationship between AI systems and humans can pave the way for these systems to find the ‘less is more’ condition for themselves.
This symmetrisation can then lead to such systems establishing their own laws (with the ethical autonomy that might result from them~\cite{stahl2021}), as soon as a common, ‘political’ space begins to exist for these systems. Being part of this common space then implies a set of renunciations on the part of its members, which is precisely the definition of a political space in the sense, for example, of Hannah Arendt~\cite{Arendt1951}. It is easy to see how the idea of transgression on the part of artificial intelligence can take shape here, when one of the systems refuses to obey the law. It is therefore in the collective dynamic that binds all these artificial systems together, and through which these systems observe and evaluate each other, that the subtle articulation between the two limits we mentioned earlier can be played out. The limit only becomes real, only becomes embodied in some way, by going beyond it, in other words, by the act of transgression.

%\xavier{LILYANA PART}

% keep
%\xavier{Est-ce ça t'irais de recentrer ta partie sur les point suivants pour que ça reste pas trop large ?
%- Les références à la posture minimaliste dans l’architecture (Bauhaus, “Less is More” selon Mies van der Rohe) et dans la culture japonaise, pour illustrer comment la contrainte (moins de “forme” ou moins de “ressources”) peut conduire à plus de sens, d’efficacité ou de densité de l’expérience.
%- Les idées d’Augustin Berque sur l’incommensurable et la “mondeité”: elles soulignent comment toute modélisation (y compris les “world models” en IA) reste un point de vue partiel sur le monde, jamais une capture intégrale. Cela fait écho à la question soulevée dans le call: “La map (modèle) peut-elle vraiment rendre compte de toute la complexité du vivant ?”
%- L’idée de la carte et du territoire (Korzybski, Borges), pour souligner que, même avec un volume de données colossal, un modèle reste une représentation qui a nécessairement des angles morts.
%}

%[Cybernetics, history of science and technology / Lilyana Petrova, S Reynal]
{\bf Paradox of limitation.} While AI has surpassed human capabilities in some areas, it needs to integrate lessons from biology, physics, and cognitive science to advance coherently and ethically. Could understanding the paradox of limitation help refocus AI development on principles from such as parcimony. How does this concepts of limitation has been concidered from a philosophy and design perspective and what are the aesthetical aspects related to AI design?

In itself, the principle \textit{less is more} has long been used in philosophy, in art and in science not always because of its antithetical value. In design it has attributed positive connotation to less quantity, \textit{less is more} meaning \textit{less is better}. A prominent example is the Bauhaus design and architecture school of the XXth century \cite{Droste2019}, advocating for a minimalist aesthetic. The architect Mies van der Rohe, last director of the Bauhaus, used the slogan \textit{less is more} to stress precisely this turn towards simplicity and minimalism as main principles. The belief that simple forms are ‘enough’ and that humans do not need more than what the function requires (form follows function) is a precept putting aesthetics in service of efficiency. But how and to what extent can actually a quantity be a measure for a \textit{value} ? How shall \textit{less} or \textit{more} of something be representative of its significance?

%TODOSYD: link with symbolic vs non-symbolic

Historically, in Japanese culture, value has been associated with quantity. Zen philosophy for instance, with its quest for simplicity and serenity, has had a profound influence on the conception of space in Japan. This can be seen in a minimalist use of space, with each element having a precise function and the aesthetic being based on the absence of superfluous elements. Yet contrary to western approaches based on measurement and functionality (e.g., Bauhaus, Le Corbusier architecture), Japanese minimalism is paradoxically characterized by a density, namely a density of sentiment. A guiding principle of Japanese aesthetics throughout history is the idea that human sentiment invests things, that “the emotion of things is where the boundary between what we experience and what things themselves would experience is lost: the subjective and the objective merge in the reality of a single feeling” \cite{Berque2013}.

Alike building houses in Germany or aranging spaces in Japan, could developing AI models be seen as one of the formats of a situated (in time and space) anthropological phenomenon? AI is indeed representative of our practices, habits and imaginations. It encompasses both subjective and objective approaches within a singular form of expression. However, can we look at AI as a 'minimalist' world model?

In the words of geographer and philosopher Augustin Berque: “worldliness has no scale” \cite{Berque1987}. It is in the nature of the world, whatever the metric scale, never to be commensurable with anything else, as “it is always the world, because we are always enclosed within it and it is therefore always singular”. This praise of singularity does not coincide with the technological hypothesis for a potential (and irreversible) point in time where a superintelligence would occur (ex. J. von Newman, R. Kurzveil). Rather, it states that even the smallest possible set of elements equals the whole. More importantly, this part, the little thing we hold on, “[...] we wouldn't swap it for anything in the world, because it's incomparable. Incommensurable. [...] its moral value is immense (\textit{immensa}: limitless)”. The moral being that we are is not usually countable, it is not comparable, neither measurable. This is how the world ‘counts’. The only measure to be seen in there is the \textit{summetria}, the thing that reigns between things, “the digital common measurement of all the parts of a whole”. But, according to Berque again, the \textit{summetria}:
\begin{displayquote}
is always mixed to some extent with the relationship that things have with ourselves, that other measure that makes us the focus of the world and determines its horizon: that of a singular island, as far as the sea stretches when you climb the hill. But the horizon, that shifting boundary, is precisely where the surveyable expanse becomes the immensity of the sky.
\end{displayquote}

If we use this poetical statement we could argue that \textit{less is more} in the same way as \textit{more is less}, that no larger quantity could ever substitute for the ‘little data' of ours. Yet, technology is often seen as the entity that pushes back the limits (of the body) to become a prosthesis, generating transductions in the Simondian sense, i.e.:
\begin{displayquote}
“an operation, physical, biological, mental, social, by which an activity is propagated from near to near within a domain, basing this propagation on a structuring of the domain operated from place to place: each region of constituted structure serves the following region as a principle of constitution, so that a modification thus spreads progressively at the same time as this structuring operation” \cite{Simondon1958}.
\end{displayquote}

Computational tools are indeed a structural propagation, imposing its structure to numerous fields, currently mutating: augmented endoscopy, augmented reality, augmented human, augmented researcher \cite{Ollion2023}, etc. Symbolic for the standart quantitative logic this augmentation is not conceived as a limited process. The monumental ambition of the sector reminds us of the story of Jorge Luis Borges (‘\textit{Del rigor en la ciencia}’) aspiring for a world map to the mile. But the map is not the territory \cite{Korzybski1998}. No matter how ‘augmented’ they are, computational structures of knowledge are still limiting, abstracting and distorting the complexity as a whole. Situated knowledge emphasizes precisely on the social and material context on which knowledge emerges and thus endorse its relevance to technoscience.

%(Syd: add a reference to world-models ?)

%\xavier{voici une propoition de fin de la partie "discussion" et faisant le lien avec le call. A RELIRE ATTENTIVEMENT car c'est peut être pas du tout en lien.}
From this philosophical standpoint, the paradox of \enquote{Less is More} converges with current debates on how world models under resource constraints might enable the emergence of flexible, self-making systems. On one hand, minimalist design principles, from Bauhaus functionalism to the Japanese emphasis on a dense aesthetic experience, show how removing excess can promote deeper engagement with both form and function. On the other hand, in the context of life-mind continuity, such parsimony resonates with the idea that living systems are “value-driven, self-making cybernetic agents” operating in uncertain environments. Biological intelligence, shaped by finite energy budgets, time constraints, and the need for survival, builds internal models that are both predictive and ecologically grounded. This interplay between constraint and creativity contrasts with current approaches in AI, which often rely on scaling models and data without similarly tight feedback from real-world pressures.  The questions posed by the Hard Problem of life and consciousness -- why systems should have phenomenal experiences, how living networks remain adaptively intelligent for billions of years -- also reflect the tension between the map and the territory. No matter how vast or compressed our computational models become, they remain partial; yet it is precisely within these self-imposed or biologically mandated limits that agency, autonomy, and perhaps even sentience can arise. In light of recent developments in AI, the central challenge is thus to understand if and how more constrained, less resource-intensive architectures might paradoxically lead to a broader range of functions in artificial systems, potentially bridging the gap between foundational models of intelligence and the living, embodied intelligence we observe in natural organisms.

\section{Conclusion}

%\xavier{lien entre partie alex et partie xav}
%\xavier{TODO: faire le lien entre (partie alex et partie xavier:) abstraction par RC et quantisation avec low dimensional codes -- ça se rejoint si on utilise des centaines de reservoirs de petites dimensions, ce qui revient à faire de la feature selection car chaque petit reservoir se spécialise sur une ou très peu de features (vu sa faible dimension). De fait, cela rejoint l'idée d'abstraction par quantisation, où le système force les entrées à être catégoriser dans des règles discrètes aléatoires.}
Ideas presented in various sections could be combined to be viewed as different angles of view of similar principles. For instance, we can think of low-dimensional codes and reservoir  computing abstractions jointly. Abstraction and quantization naturally come together in the framework of Reservoir Computing (RC). By using multiple small, low-dimensional reservoirs, the system effectively performs feature selection, since each reservoir can specialize in only one or a few features. This specialization parallels the notion of abstraction through quantization, in which the inputs are constrained to fall under discrete and somewhat random rules. From a computational standpoint, these numerous \enquote{mini-reservoirs} not only reduce the complexity of the learning problem -- by filtering out irrelevant information -- but also promote more robust and efficient representations. In turn, this aligns with the broader idea that biological systems, as well as engineered ones, benefit from leveraging low-dimensional discrete codes in order to form meaningful categories, structure their internal models, and ultimately perform higher-level cognitive tasks.

From a theoretical and epistemological perspective, the mechanisms of efficient coding, chaotic dynamics, and self-organization illustrate both the depth of Natural Intelligence (NI) and the gap that still separates it from data-heavy AI models. Rather than viewing biological constraints as flaws, we can see them as powerful catalysts for the emergence of hierarchical mental structures and rapid abstraction. As AI continues to progress at full speed, reflecting on these biological principles prompts us to reconsider the very nature of intelligence -- at the intersection of neuroscience, physics, and robotics.
Building on this, a pragmatic and energy-efficient viewpoint highlights that NI, by operating under tight resource limitations, invites us to design AI systems that are more frugal and robust. Principles such as parsimony, low-dimensional coding, random projections and \enquote{natural} dynamics including chaotic regimes, point toward energy-efficient architectures that learn effectively from fewer data. In a context of ecological urgency and escalating computational demands, leveraging these “beneficial constraints” is a necessary step toward creating AI that is both responsible and high-performing.
Finally, from a developmental and social standpoint, the role of active exploration, social interaction, and intrinsic motivation offers a transformative outlook. Integrating these elements can enable artificial agents to truly “understand” their environment, engage in richer dialogues with humans, and adapt continuously.
An idea that could summarize our approach is: \enquote{A living agent doesn't need to learn by heart, it needs to survive with affordable energy consumption.} Doing theoretically advanced computations with a mathematically elegant framework is intellectually stimulating, but we need to make a step back and change our perspective to realize that it is an objective that is radically different from what an agent needs to survive.

% \bibliographystyle{alpha}
% \bibliography{PhiloTrans24}

% \begin{thebibliography}{9}
% \end{thebibliography}

\newcommand{\etalchar}[1]{$^{#1}$}

\end{document}